\documentclass[prd,aps,floatfix,superscriptaddress,nofootinbib,twocolumn,10pt,English]{revtex4}

\usepackage{amssymb,amsmath}
\usepackage{graphicx}
\usepackage[lofdepth,lotdepth]{subfig}
\usepackage[percent]{overpic}
\usepackage{overpic}
\usepackage{color}
\usepackage[colorlinks=true]{hyperref}
\numberwithin{equation}{section}

\begin{document}

\title{Relating Black Hole Shadow to Quasinormal Modes for Rotating Black Holes}

\author{Huan Yang}
\affiliation{Perimeter Institute for Theoretical Physics, Waterloo, Ontario N2L 2Y5, Canada}
\affiliation{University of Guelph, Guelph, Ontario N1G 2W1, Canada}

\begin{abstract}
In this work, we explore the connection between the critical curves (``shadows")  and the  quasinormal mode frequencies (in the eikonal limit) of Kerr black holes. This mapping has been previously established for non-rotating black holes. We show that, the shadow seen by an distant observer at a given inclination angle, can be mapped to a family of quasinormal modes with  $m/(\ell+1/2)$ bounded within certain range, where $m$ is is azimuthal node number and $\ell$ is the angular node number. We discuss the possibility of testing such relation with space-borne gravitational wave detectors and the next-generation Event Horizon Telescope.\end{abstract}

\maketitle

\section{Introduction}\label{sec:1}
\label{sec:intro}

The Event Horizon Telescope (EHT) Collaboration has annouced the first black hole image (M87) in 2019 thanks to advancing techniques in radio interferometry \cite{Akiyama:2019cqa}. It is expected that Sgr A* is the next viable source for black hoe imaging, because it has comparable angular diameter in the sky as M87.
Moving forward, it is interesting to consider science opportunities asscociated with next-generation EHT-like observatories, both in the front of constraining astrophysics models and in the aspect of testing the theory of General Relativity.

In addition to possible gravity test with M87 data \cite{Psaltis:2020lvx}, which generated debates about its validity \cite{Volkel:2020xlc,Gralla:2020pra}, there are already discussions on resolving novel signatures of black hole photon ring, assuming future space-based  telescopes for radio interferometry.
As shown in \cite{johnson:2019ljv, Gralla:2020nwp}, the size and shape of the black hole critical curve, or more precisely the photon orbit, are encode in the visibility function. This potentially enables percent-level test on the prediction of General Relativity in the strong-gravity regime with space-based radio interferometry \cite{Gralla:2020srx}.
On the other hand, gravitaitonal wave (GW) observation of binary black hole merger by ground-based and space-borne GW detectors provide another route for strong-gravity tests of General Relativity \cite{Berti:2018vdi,Berti:2018cxi}. It is therefore very interesting to compare these two sets of observables, if both are available for certain blakc hole systems. 

Considering a black hole which is non-rotating, it is shown that \cite{Stefanov:2010xz,Jusufi:2019ltj} the observables from radio imaging of the black hole shadow and gravitational wave measurement of the black hole quasirnormal mode (QNM) frequency ($\omega_{\ell m n} = \omega_{\rm R} -i \omega_{\rm I}$) are related by the following simple formulas: 
\begin{align}\label{eq:s}
R & = {\rm lim}_{\ell \rightarrow \infty}\, \frac{\ell}{\omega_{\rm R}}\,,\nonumber \\
\frac{\log \tilde{r}}{2 \pi R} & =  {\rm lim}_{\ell \rightarrow \infty}\,\omega_{\rm I} \,,
\end{align}
where $R$ is the radius of the black hole shadow, $\tilde{r}$ is the amplitude ratio between the $N$th image and the $(N+2)$th image \footnote{Notice that in the original statement of \cite{Stefanov:2010xz,Jusufi:2019ltj} $\tilde{r}$ represents the flux ratio instead of amplitude ratio, where they compare successive images.}.  The $\ell \rightarrow \infty$ limit applied here is usually referred as the eikonal limit. The relations described in Eq.~\ref{eq:s} not only hold for Schwarzschld spacetime, but also for general spherically symmetric spacetimes \cite{Stefanov:2010xz,Jusufi:2019ltj}. Intuitively speaking, both the black hole critical curve (shadow) and the black hole quasinormal modes are related to the geometric signatures of the photon ring \cite{Cardoso:2008bp}, so that they must be connected. However, an explicit relation as Eq.~\ref{eq:s} still hints for a physical interpretation. In Sec. \ref{sec:r} we present a simple derivation for this relation using the Hamilton's principal function.

Massive black holes in galactic centres are likely rotating due to their accretion history and/or possible merger history. Therefore it is astrophysically relevant to check whether the analogy of  Eq.~\ref{eq:s} exist for general rotating black holes. By using the Hamilton's principal function for geometric rays and applying the geometric optic correspondence for Kerr quasinormal modes \cite{Yang:2012he}, we are able to derive the mapping between the Kerr critical curves, which are no longer circular, and the Kerr quasinormal modes in the eikonal limit.
This finding completes the picture for the mapping between these two sets of observables for astrophysical black holes.

Observationally it is challenging to find a massive black hole system which simultaneously allow horizon-scale radio interferometry and gravitational wave ringdown measurement. As pointed out in \cite{1842463}, a significant fraction of extreme mass-ratio insprirals (EMRIs) in the LISA band (Laser Interferometry Space Antenna) should be accompanied by Active Galactic Nuclei (AGN) because accretion disks dramatically increase the EMRI formation rate.  However, EMRIs observed by LISA tend to have rather faint ringdown signal \cite{baibhav2019probing}, which is not ideal for black hole spectroscopy measurements. On the other hand, massive black hole mergers could produce loud ringdown signals for LISA detection \cite{Berti:2005ys}. If such system is embedded in a gas-rich environment, the followon radio interferometry observation may further reveal details of the final black hole shadow. The main constraint, however, comes from the fact that these sources are usually located at cosmological distances. Therefore the angular resolution of the radio telescope has to be extremely sensitive to resolve the horizon scale image of the final black hole.

In the following we first derive the mapping between black hole shadow and eikonal quasinormal modes for generic spherically symmetric spacetimes and the Kerr spacetime. After that we discuss how to formulate the test for Kerr black holes,  the associated accuracy and what are the observational requirements for such test. 

\section{Relation between black hole QNM and shadow}\label{sec:r}

In this section, we first briefly review the geometric interpretation of black hole quasinormal modes in the eikonal limit.
After that we present a simple derivation for Eq.~\ref{eq:s} for general spherically symmetric spacetime. Using similar approach we also derive the mapping relation in the case of rotating black holes.

\subsection{Quasinormal modes in the eikonal limit}
In the eikonal limit, the wavelength  is much shorter than the size of the black hole. As a result, the (possible) polarization of the wave is a subdominant factor for wave propagation, so that gravitational/electromagnetic/scalar quasinormal modes all share the same leading order frequency formula. For Schwarzschild black hole \cite{Cardoso:2008bp} and Kerr black hole \cite{Yang:2012he}, it has been shown that QNMs can be mapping to geometric rays propagating along circular/spherical photon orbits, and their frequencies are related to the geometric properties of the photon orbits. In particular,  
we have
\begin{align}
L_z & \leftrightarrow m, \quad
E  \leftrightarrow \omega_{\rm R} \nonumber \\
\mathcal{D} +L^2_z &\leftrightarrow {\rm Re}(A_{\ell m}) ,\quad
\gamma \leftrightarrow \omega_{\rm I}\,,
\end{align}
where $E, L_z, \mathcal{D}, \gamma$ are the energy, azimuthal angular momentum, Carter's constant and the Lyapunov exponent of the geometric ray moving along spherical photon orbits, and $A_{\ell m}$ is the angular eigenvalue of the associated quasinormal mode, obtained by solving the angular Teukolsky equation.
The geometric correspondence also enables one to derive the eikonal  QNM frequency formula in terms of the angular and precession frequencies $\Omega_{\rm \theta}, \Omega_{\rm prec}$ of the null ray \cite{Yang:2012he} (also see Eq.~\ref{eq:do}):
\begin{align}\label{eq:e2}
\omega_{\rm \ell m n} \approx \left ( \ell+\frac{1}{2}\right ) \Omega_{\rm R}(\mu) - i \left ( n+\frac{1}{2}\right ) \Omega_{\rm I}(\mu)
\end{align}
with $\mu \equiv m/(l+1/2)$ and
\begin{align}
\Omega_{\rm R} = \Omega_{\rm \theta}(\mu) + \mu \Omega_{\rm prec}(\mu),\quad \Omega_{\rm I} =\gamma\,.
\end{align}
In principle by including all higher-order correction as powers of $\mathcal{O}(1/(\ell+1/2))$, the quasinormal mode frequency $\omega_{\rm \ell m n}$ can be fully recovered as a summation of the power series, with coefficients being functions of $\mu$. In practice, the relative error of  in Eq.~\ref{eq:e2} roughly scales as $1/(\ell+1/2)^2$, which is already less than $10\%$ accuracy for $\ell \ge 3$.

\subsection{Spherically symmetric spacetimes}

Consider a null ray propagating in a general spherically symmetric spacetime. Because of the spherical symmetry, we can always choose the coordinate system such that the null rays only move on the equatorial plane. In other words, the null ray has two degrees of freedom and two conserved quantities: energy and angular momentum (along the axis perpendicular to the plane.) This also means that the motion has to be separable. In the language of Hamilton-Jacobi method, the principal function can be written as
\begin{align}
S = S_r(r) +L_z \phi- E t\,,
\end{align}
which has the physical meaning of the phase of the null ray. It has to be invariant along the propagation for the null ray, and in particular for the one moves along the circular photon orbit, we have
\begin{align}
dS = L_z d \phi - E dt =0
\end{align}
where $d S_r$ is zero because of the absence of radial motion. We notice that $L_z/E$ is exactly the radius $R$ of the critical curve, considering any photon orbit infinitesimally deviate from the circular photon orbit and eventually escape to infinity; $\Omega_\phi =d \phi/d t$ is the same as $\omega_{\rm R}/\ell$ in the eikonal limit due to the geometric optic correspondence, so that the first line of Eq.~\ref{eq:s} can be recovered. On the other hand, going beyond the leading order WKB analysis \cite{Yang:2012he}, it can be shown that the amplitude of the wave decays as $e^{-\gamma t}$. After one orbital cycle $T=2\pi/\Omega_\phi = 2 \pi  R$, the amplitude has decayed an additional factor $e^{-\gamma T} =e^{-\gamma 2 \pi R}$. Therefore the amplitude ratio between the images is indeed described by the second line of Eq.~\ref{eq:s}.

In \cite{Jusufi:2020dhz} the relation between Kerr quasinormal modes and shadow was analyzed for equatorial photon orbits and $m =\pm \ell$ modes. The method outlines in this section naturally applies in this scenario. In the next section we discuss the mapping for generic Kerr quasinormal modes, which are related to general spherical photon orbits.

\subsection{Kerr spacetime}

The motion of geometric rays in Kerr spacetime is also separable, thanks to the nontrivial conserved quantity $\mathcal{D}$: the Carter's constant, in addition to $E$ and $L_z$. Its principal function can be written as
\begin{align}
S &= S_r(r) +S_\theta(\theta)+L_z \phi- E t\,, \nonumber \\
& = \pm \int dr \sqrt{\mathcal{R}} \pm \int d\theta \sqrt{\Theta} + L_z \phi - E t
\end{align}
where
\begin{align}
\mathcal{R} & =  [E(r^2+a^2)-L_z a]^2 -\Delta [(L_z-a E)^2+\mathcal{D}]\,, \nonumber \\
\Theta & = \mathcal{D} -\cos^2\theta (L_z^2/\sin^2\theta -a^2 E^2)\,,
\end{align}
with $\Delta = r^2 +a^2 - 2 M r$ and $a$ being the black hole spin. The $\pm$ signs for $S_r, S_\theta$ depend on the propagating direction of the geometric ray. For a null ray moving along a spherical photon orbit, the radius is constant so that we can neglect the radial term $S_r$ in the principal function, so that the sim of the rest three terms should be invariant along the propagation. Consider a full cycle in the $\theta$ direction, we have
\begin{align}\label{eq:ds}
\Delta S = 0 = \oint \sqrt{\Theta} d \theta + L_z \Delta \phi - E T_\theta\,,
\end{align}
where $T_\theta$ is the period of motion in $\theta$ direction, and $\Delta \phi$ is the azimuthal angle changed after completing a cycle in $\theta$ direction. 
Notice that $\Delta \phi$ is not the same as the {\it precession angle} $\Delta \phi_{\rm prec}$. In fact, they are related to each other by
\begin{align}\label{eq:prec}
\Delta \phi = \Delta \phi_{\rm prec} +2 \pi {\rm sgn}(L_z)
\end{align}
where ${\rm sgn}( .)$ evaluates the sign of the argument.  This is easy to understand.
If we take the $a \rightarrow 0$ limit, the black holes becomes non-rotating and the photon orbit becomes circular. In this case, the periods of motion in both $\theta$ and $\phi$ direction become the same, i.e., no precession.  As we follow a full cycle in the $\theta$ direction, we should have moved $2 \pi$ in the $\phi$ direction, which is consistent with Eq.~\ref{eq:prec}.
Therefore for general Kerr orbits we need to subtract $2 \pi$ away from $\Delta \phi$ to obtain the precession angle $\Delta \phi_{\rm prec}$. In particular, we notice that
\begin{align}\label{eq:do}
\Omega_\theta = \frac{2 \pi}{T_\theta}, \quad \Omega_{\rm prec} = \frac{\Delta \phi_{\rm prec}}{T_\theta}\,.
\end{align}

\begin{figure}
\includegraphics[scale=0.45]{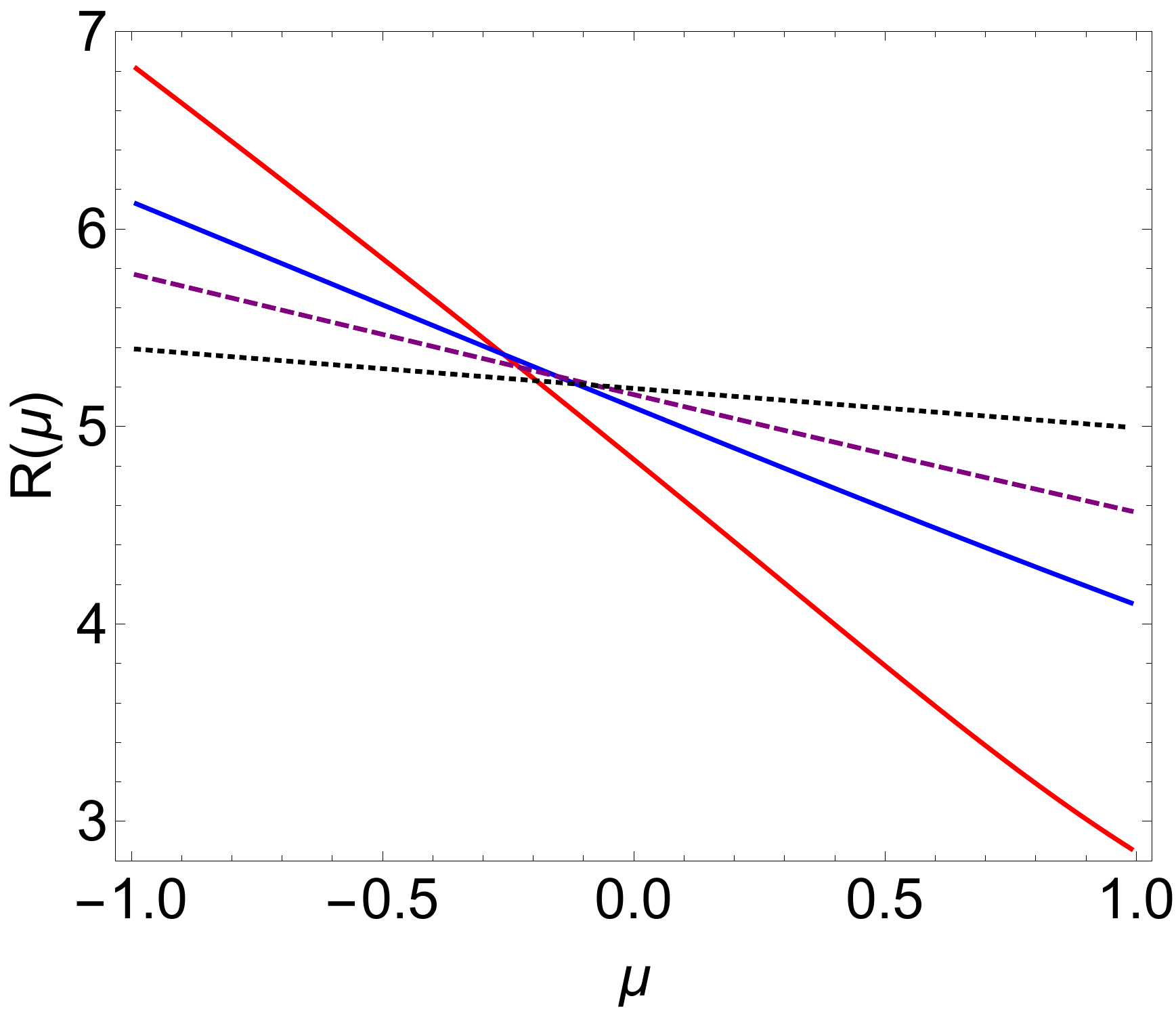}
\caption{\label{fig:1}  The impact parameter $R(\mu)$ as a function of $\mu$ for various black hole spins: $a/M = 0.9$ (Solid Red), $a/M = 0.5$ (Solid Blue), $a/M = 0.3$ (Dashed Purple), $a/M = 0.1$ (Dotted Black).} 
\end{figure}

\begin{figure}
\includegraphics[scale=0.5]{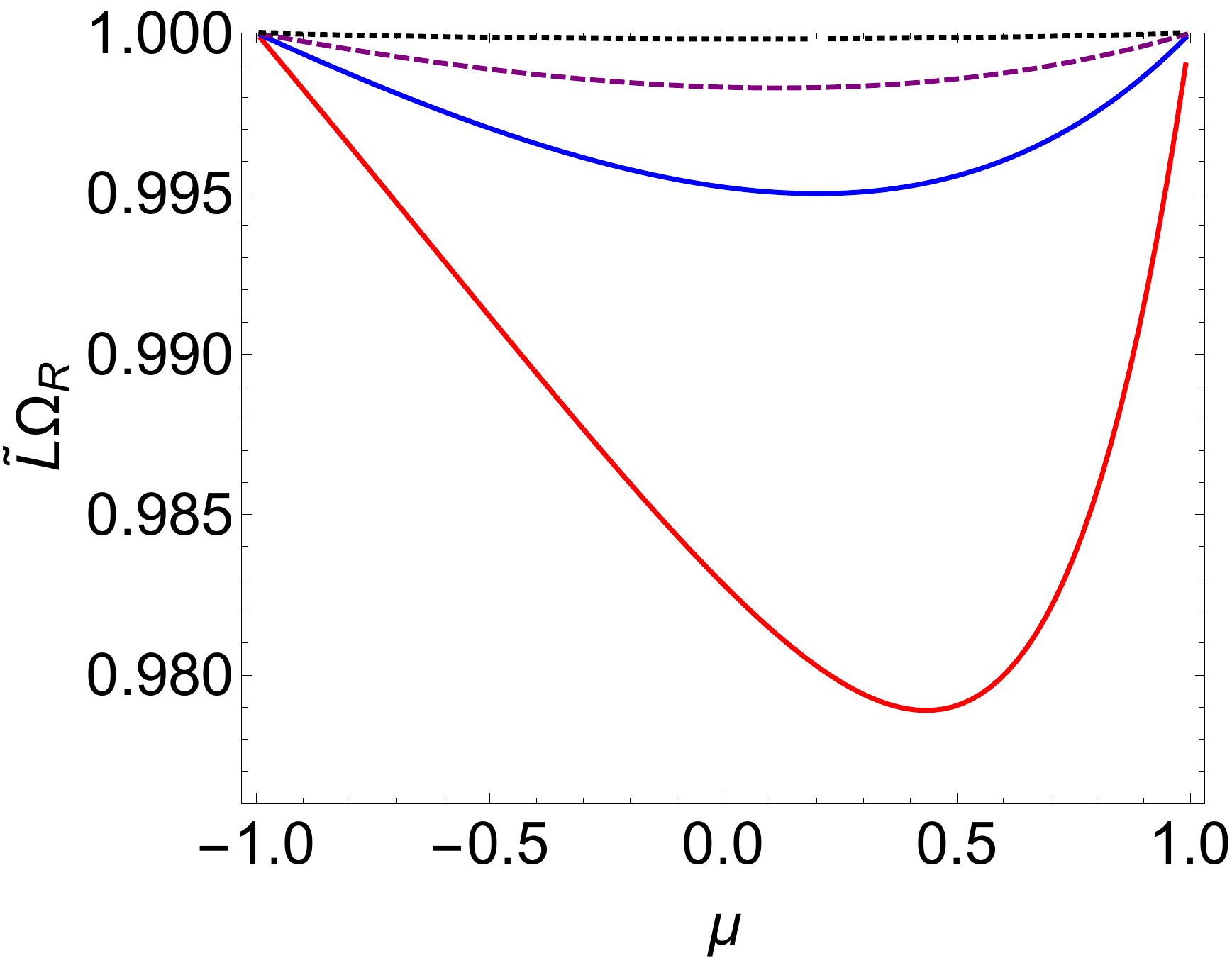}
\caption{\label{fig:2}  The product $\tilde{L} \Omega_{\rm R}$ as a function of $\mu$ for various black hole spins: $a/M = 0.9$ (Solid Red), $a/M = 0.5$ (Solid Blue), $a/M = 0.3$ (Dashed Purple), $a/M = 0.1$ (Dotted Black).} 
\end{figure}

The cycle integration in the $\theta$ direction is 
\begin{align}
\oint \sqrt{\Theta} d \theta =2 \int^{\theta_+}_{\theta_-} \sqrt{\Theta}\,,
\end{align}
where $\theta_{\pm}$ are the critical angles (assuming $\theta_+ >\theta_-$) such that $\Theta (\theta_{\pm}) =0$. Motivated by the WKB analysis for Kerr quasinormal modes in \cite{Yang:2012he}, we can rewrite this cycle integration as
\begin{align}\label{eq:bs}
2 \int^{\theta_+}_{\theta_-} \sqrt{\Theta} = 2 \pi (L - |L_z|)
\end{align}
which can be viewed as the defining equation for $L$, as a function of $E, L_z, \mathcal{D}$. Eq.~\ref{eq:bs} also has the physical meaning of the {\it Bohr-Sommerfeld Condition} as we compare with the eigenvalue problem in the $\theta$ direction for Kerr quasinormal modes \cite{Yang:2012he}. 
The expression of $L$ is not known in closed form for general cases, but there are analytical power-law expansions. In particular, for Schwarzschild black holes, it can be shown that 
\begin{align}
L = \sqrt{\mathcal{D}+L_z^2}\,, \quad (a=0)
\end{align}
i.e., $L$ is the total angular momentum. For rotating black holes, the solution of Eq.~\ref{eq:bs} reads
\begin{align}
\mathcal{D} +L_z^2 = L^2 -\frac{a^2 E^2}{2}\left ( 1-\frac{L^2_z}{L^2}\right ) +\mathcal{O}\left( \frac{a^4 E^4}{L^4}\right)
\end{align}
where the higher order terms can be found in \cite{Berti:2005gp} by replacing $\ell+1/2$ by $L$ therein and taking the eikonal limit. In addition, in the eikonal limit, we shall not distinguish $L_z/L$ for the geometric ray or $m/(l+1/2)$ for the mode \cite{Yang:2012he}, which are both referred as $\mu$. Since $\sqrt{(\mathcal{D}+L^2_z)}/E = R$ is the impact parameter of the null ray if it escapes to infinity, the above equation can be rewritten as
\begin{align}\label{eq:app}
\frac{L^2}{E^2} \approx \frac{\tilde{L}^2}{E^2} := R^2 +\frac{a^2}{2}(1-\mu^2)\,.
\end{align}
The relation between $R$ and $\mu$ can be obtained by combining Eq.~\ref{eq:bs} with Eq.~(4.1a) and Eq.~(4.1b) in \cite{Yang:2012he} \footnote{More specifically, for any viable radius of the spherical photon orbits, we can first use Eq.~(4.1a) and Eq.~(4.1b)  to find $\mathcal{D}/E^2$ and $L_z/E$, and then use Eq.~\ref{eq:bs} to find the corresponding $L$ and $\mu$. It is then straightforward to obtain $R(\mu)$.}. In Fig.~\ref{fig:1} we illustrate its function dependence for various black hole spins: $a/M = 0.1, 0.3, 0.5, 0.9$.

Now Eq.~\ref{eq:bs}, together with Eq.~\ref{eq:ds} suggest that
\begin{align}
\frac{2 \pi L}{E} +L_z \frac{\Delta \phi_{\rm prec}}{E} -T_\theta =0\,,
\end{align}
which implies that
\begin{align}\label{eq:real}
\frac{L}{E} = \frac{1}{\Omega_\theta +\mu \Omega_{\rm prec}} = \frac{1}{\Omega_R}
\end{align}
as a generalization of Eq.~\ref{eq:s} for Kerr black holes. We may use Eq.~\ref{eq:app} as an approximation for $L/E$ to connect the radius of the critical curve to the quasinormal mode frequency. In Fig.~\ref{fig:2}, we plot $\tilde{L} \Omega_{\rm R}$ as an approximation for $L \Omega_{\rm R} =1$ for various black hole spin $a$ and different $\mu$. We find that Eq.~\ref{eq:app} gives a reasonable approximation with error at the percent level. Therefore it should suffice for testing General Relativity with uncertainty at the percent level.

For any point source near the rotating black hole, its wave emission as received by a distant observer, may be evaluated using the Kerr Green's function \cite{Yang:2013shb}.
The lensed images can be classified into two categories, with even and odd winding numbers respectively, and they arrive at the observer in alternating order. For high order images, i.e., the observation time $T$ is large, the phase factor of the wave reads \cite{Yang:2013shb}
\begin{align}\label{eq:phase1}
g(\mu) \approx &-\Omega_{\rm R} T+\mu (\pi W +\phi -\phi') \pm \frac{\Phi(\theta,\theta')}{L} \nonumber \\
&+\bar{\alpha}_1(r) +\bar{\alpha}_1(r')\,,
\end{align}
where $W \in \mathcal{Z}$ is another set of winding number associated with ray precession near the black hole (it is zero for Schwarzschild black holes), $(r,\theta,\phi)$ and $(r',\theta',\phi')$ are the coordinate positions of the emitter and the receiver respectively. The definitions of $\Phi, \bar{\alpha}_1$ can be found in Sec. V in \cite{Yang:2013shb}, which are not essential for the discussion here. For given $T$ and emitter/receiver locations, the ray geometric phase $g({\mu_0})$ is evaluated with $\mu_0$ satisfying $g'(\mu) |_{\mu=\mu_0}=0$.

Let us now denote $T_0$ and $T_0+\Delta T$ as the arriving times of successive even images or those of the odd images, with $\mu_0$ and $\mu_0+\Delta \mu_0$ respectively. Eq.~\ref{eq:phase1} implies that
\begin{align}\label{eq:d1}
 \Delta (\Omega'_{R} T) = \Delta T_0 \Omega'_{\rm R} + T_0 \Omega''_{\rm R} \Delta \mu_0 =0\,.
\end{align}
In addition, since the phase difference between successive even/odd images is $g(\mu_0+\Delta \mu_0)-g(\mu_0) = 2 \pi$ \cite{Yang:2013shb}, we have
\begin{align}\label{eq:d2}
\Omega_{\rm R} \Delta T_0 + \Omega'_{\rm R} \Delta \mu_0 T_0 =2 \pi\,.
\end{align}
At last, we notice that the amplitude ratio between these two successive even (odd) images, or the ratio $\tilde{r}_{N,N+2}$ between the $N$th and the $N+2$th image is (combining Eq.~\ref{eq:d1} and Eq.~\ref{eq:d2})
\begin{align}
\log \tilde{r}_{N,N+2} &=\Omega_{\rm I}(\mu_0 +\Delta \mu_0)( T_0+\Delta T_0)-\Omega_{\rm I}(\mu_0) T_0 \nonumber \\
& \approx \Omega'_{\rm I} \Delta \mu_0 T_0 +\Omega_{\rm I } \Delta T_0 \nonumber \\
& = 2 \pi \frac{\Omega_{\rm I} - \Omega'_{\rm I} \Omega'_{\rm R}/\Omega''_{\rm R}}{\Omega_{\rm R} -(\Omega'_{\rm R})^2/\Omega''_{\rm R}}\,.
\end{align}
For Schwarzschild black hole $\Delta \mu_0 =0$ so that the amplitude ratio reduces to $2 \pi \Omega_{\rm I}/\Omega_{\rm R}$.

\subsection{Possible test}\label{sec:pt}

\begin{figure}
\includegraphics[scale=0.45]{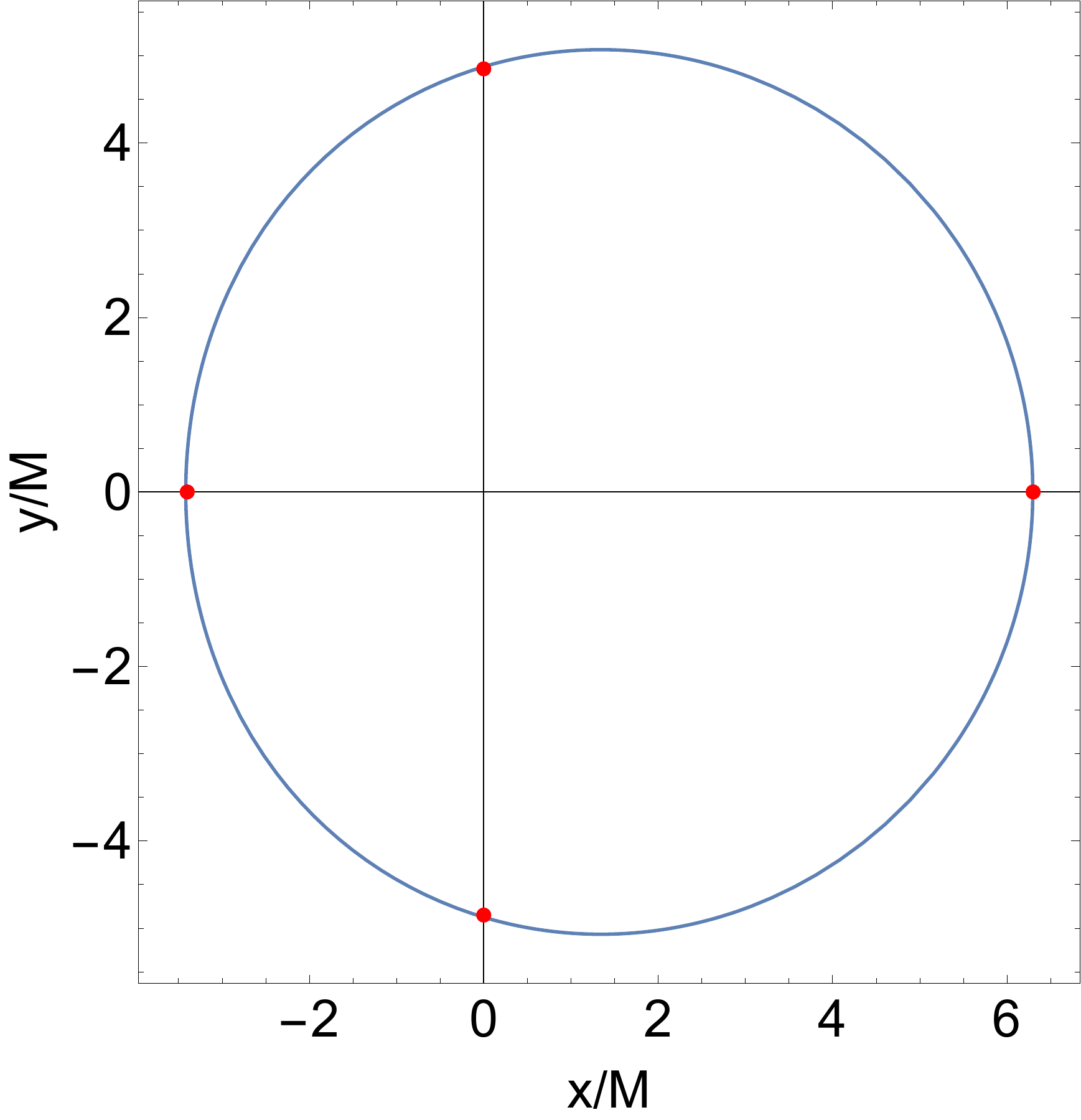}
\caption{\label{fig:3}  The critical curve of a Kerr black hole with $a/M=0.9$ and $\iota=\pi/4$. The red dots on the y axis represent the points with $\mu =0$. The left/right dots on the x axis correspond to the maximum/minimum $\mu = 0.683/-0.725$. For other points in the critical curve, the corresponding $\mu$ can be found using its radius and Fig.~\ref{fig:1}.} 
\end{figure}

Since both the gravitational wave observable and the radio interferometry observable can be used to separately measure/constrain the black hole mass $M$, spin $a$ and the inclination angle $\iota$ of the observer, one can directly compare these parameters as inferred by GW and EM observation separately to test General Relativity, in the same spirit as the consistency test comparing parameters extracted  from inspiral/ringdown waveform \cite{TheLIGOScientific:2016src}. On the other hand, with the relation between the shadow size and quasinormal mode frequency depicted in Eq.~\ref{eq:real}, it becomes possible to directly  use these length/time observables to check the validity of Eq.~\ref{eq:real} for gravity tests.

Depending on the black hole spin and the inclination angle, the shape of the critical curve varies. For example, in Fig.~\ref{fig:3}, we present the critical curve of a Kerr black hole with $a/M =0.9$, where the observer is located with $\iota = \pi/4$. Each point on the critical curve corresponds to a spherical photon orbit with certain $\mu$. In order to find the associated $\mu$, the easiest way is to first evaluate its impact radius $R$ and then invert the $R(\mu)$ function  shown in Fig.~\ref{fig:1}. In particular, we notice that the two points on the y-axis have $\mu = 0$, corresponding to the photon orbit with $L_z =0$.

Given a critical curve, it is still nontrivial to test Eq.~\ref{eq:real} because (i) it is difficult to determine the $\mu$s of target points on the curve without using the timescale information from  gravitational wave measurements, and (ii) a discrete set of QNMs measured  from black hole spectroscopy usually sample a discrete set of $\mu$s, which may not cover the target points on the critical curve. For example, if only the $\ell=2, m=2$  and $\ell=3, m=3$ mode are measured, we can at best directly constrain the cases with $\mu=0.8, 6/7$, which are outside the range of $\mu$ available in Fig.~\ref{fig:3}.

The points associated with $\mu=0$ better serve the purpose for such test. Assuming $M, a, \iota$ can be measured by gravitational wave observation, one can compute $R(\mu_{\rm max})/R(\mu_{\rm min})$ to identify the centre of the critical curve. Notice that we have used General Relativity and the values of $a/M, \iota$ in this step,  but 
not the absolute timescale information from the gravitational wave observation. Therefore Eq.~\ref{eq:real} can be tested for the special case
\begin{align}\label{eq:mu0}
\sqrt{R(\mu=0)^2+a^2/2} \approx \frac{1}{\Omega_{\rm R}(\mu=0)}\,,
\end{align}
where $\Omega(\mu=0) \approx  \mathcal{R}(\omega_{\ell 0 0})/(\ell+1/2)$ can be estimated from the observation of $m=0$ modes. According to Fig.~ 6 of \cite{Yang:2012he}, the relative error  between $ \mathcal{R}(\omega_{\ell 0 0})/$ and $(\ell+1/2)\Omega_{\rm R}$ roughly scales as $1.2 L^{-2}$ which is below $10\%$ for $\ell \ge 3$. For realistic implementation, it is probably better to first compute $(\ell+1/2)\Omega_{\rm R}(\mu=0)/\mathcal{R}(\omega_{\ell 0 0})$ using the $a/M$ from the inspiral measurement, and then correct the measured ringdown frequency with this factor. 

\subsection{Detection aspect}\label{sec:da}

Recently it is suggested that AGN (Active Galactic Nuclei)-assisted EMRIs have comparable, if not higher formation rate comparing to the canonical channel with multibody scattering and gravitational capture \cite{1842463}. This means a large fraction of EMRIs observed by LISA may be accompanied by accretion disks, which are ideal for multi-messenger observations. However, most of the signal-to-noise ratio (SNR) of EMRIs is contributed by the inspiral part of the waveform, so that the ringdown SNR is small (typically $< 1$).
Therefore EMRIs are likely not applicable for the gravity test discussed in this work.

On the other hand, LISA is expected to observe $\mathcal{O}(1)$ to $\mathcal{O}(10)$ massive black hole (MBH) mergers every year \cite{Klein:2015hvg,Salcido:2016oor,Bonetti:2018tpf,Katz:2019qlu}. These MBH mergers are often the consequence of galaxy mergers, and they often produce very loud (SNR up to $\mathcal{O}(10^3)$) signal for the LISA observation. As a result, they are ideal sources for black hole spectroscopy and associated gravity tests \cite{Berti:2005ys}. If the MBH binary is embedded in a gas-rich environment,  we may observe radio emissions from the post-merger accretion as well. The limiting factor, however, is that these events are usually expected at cosmological distances. Imagine a post-merger black hole with mass $\sim 10^8 M_\odot$  located at redshift $z =1$, the angular size of its horizon is roughly $1.6 \times 10^{-4}$ times smaller than that of $M87$. This impose length requirement on the baseline of a spaced-based EHT, to be able to resolve the horizon scale emission from these systems.

Considering the future beyond LISA, there are already design and science case studies for the next generation space-borne gravitational wave detectors, such as the AMIGO (Advanced Millhertz Gravitational-wave Observatory) proposed in \cite{baibhav2019probing}. The new detector may achieve ten times sensitivity improvement across broadband as compared with LISA, so that the black hole spectroscopy measurement may be able to detect higher-order quasinormal modes and the detector can probe MBHs with larger masses. Therefore the next-generation detectors may open up new opportunities for the MBH merger scenario discussed above.

\section{Conclusion}

In this work, motivated by previous studies on mapping black hole quasinormal mode to back hole shadow for spherically symmetric spacetimes, we show that similar mapping also exist for rotating black holes, with derivation of its explicit form. On the one hand, it is theoretically interesting to observe that generic rotating black holes still allow a simple relation as Eq.~\ref{eq:real}. On the other hand, such mapping provides a direct mean to compare the time observable in gravitational wave measurements to the length observable in electromagnetic wave measurements for systems in the strong-gravity regime. It is however a challenging task as MBH mergers are usually happening at cosmological distances, so that an order of $\mathcal{O}(10^4)$ improvement in the angular resolution is required for future detectors.

There are several additional points worth to notice. First of all, even without horizon-scale resolution on the accretion details, the multi-messenger observation of a MBH binary or EMRI mentioned in Sec.~\ref{sec:da} still provide vast opportunities for studying accretion physics and building cosmic distance rulers. In particular, it has been argued \cite{kocsis2011observable,barausse2014can} that EMRIs embedded in accretion may be affected by disk forces so that the phase of their gravitational waveform can be significantly modified \footnote{Other environmental effects, such as the tidal gravitational field generated by other compact objects \cite{bonga2019tidal,Yang:2019iqa,yang2017general}, may also modify the EMRI waveform. It is therefore important to develop waveform models for these possible environmental effects.}. By measuring this gravitational phase shift,  disk properties, such as the density, may be constrained and further synthesized with data from electromagnetic observation at larger scales to better understand the accreting system. This is a direction worth future explorations given the prediction of AGN-assisted EMRI rate \cite{1842463}.

Secondly, black hole critical curves are not directly measured by EHT radio interferometry. A more observationally relevant notion may be photon orbits which reside near the critical curve \cite{Gralla:2019drh}. Moreover, astrophysical complications in the accretion system and the emission mechanisms should generate extra uncertainties in the length measurement of the image. Therefore the black hole critical curve discussed here should be viewed as a characteristic signature inferred from the black hole image.

In Sec.~\ref{sec:pt} we have focused on possible method of directly  testing Eq.~\ref{eq:real}. Alternatively, one can first compute the mass $M$ and spin $a$ of the MBH using only the gravitational wave data or only the electromagnetic observations, and then compare these two separate measurement. 
In this way, instead of applying only one point on the critical curve for the test (Eq.~\ref{eq:mu0}), all the visibility data can be used to provide  better constraints on $M$ and $a$.

A general gravity test may be performed as the consistency check of any prediction of General Relativity, which is the approach adopted in this work, or it can be carried out as a search for predicted deviation from General Relativity based on certain modified gravity theory. There are already many discussions in literature regarding the relation between quasinormal modes and black hole shadow for modified gravity theories \cite{Liu:2020ola,Konoplya:2020bxa} and/or additional matter content \cite{Jusufi:2019ltj}. It is however important to note that the geometric correspondence found for Schwarzschild/Kerr black holes may not hold for these modified black holes, partially because the interplay between the gravitational perturbation and the perturbation of additional fields in those models. In other words, in the eikonal limit, quasinormal modes in modified gravity theories may not directly correspond to frequencies of photon orbits.

\acknowledgements
H. Y. thanks Michael Johnson for the email exchange which motivated this work, and Zhen Pan for reading through the manuscript and having interesting discussions. H. Y. is supported by the Natural Sciences and
Engineering Research Council of Canada and in part by
Perimeter Institute for Theoretical Physics. Research at
Perimeter Institute is supported in part by the Government of
Canada through the Department of Innovation, Science and
Economic Development Canada and by the Province of Ontario through the Ministry of Colleges and Universities.


\bibliography{References}
\end{document}